\documentclass[sigconf]{acmart}

\AtBeginDocument{}
\setcopyright{rightsretained}
\copyrightyear{2026}
\acmYear{2026}
\acmConference[USRW '26]{First Workshop on Unified Search and Recommendation}{October 2, 2026}{Minneapolis, MN, USA}
\acmBooktitle{Proceedings of the First Workshop on Unified Search and Recommendation (USRW '26), October 2, 2026, Minneapolis, MN, USA}
\acmDOI{}
\acmISBN{}

\usepackage{amsmath}
\usepackage{array}
\usepackage{booktabs}
\usepackage{balance}
\usepackage{graphicx}
\usepackage{multirow}
\usepackage{subcaption}
\usepackage{tabularx}
\usepackage{ragged2e}

\newcolumntype{Y}{>{\RaggedRight\arraybackslash}X}
\newcolumntype{L}[1]{>{\RaggedRight\arraybackslash}p{#1}}

\newcommand{\sid}{\textsc{SID}}
\newcommand{\atc}{\textsc{ATC}}

\begin{document}

\title[One Hierarchy, Two Systems]{One Hierarchy, Two Systems: Semantic Product IDs for Discovery-Surface Ranking and Search-Page Query Reformulation}

\author{Steven Xu}
\authornote{These authors contributed equally to this research.}
\email{steven.xu@doordash.com}
\affiliation{%
  \institution{DoorDash Inc.}
  \city{San Francisco}
  \state{CA}
  \country{USA}}

\author{Sanjyot Thete}
\authornotemark[1]
\email{sanjyot.thete@doordash.com}
\affiliation{%
  \institution{DoorDash Inc.}
  \city{San Francisco}
  \state{CA}
  \country{USA}}

\author{Saathvik Dirisala}
\authornotemark[1]
\email{saathvik.dirisala@doordash.com}
\affiliation{%
  \institution{DoorDash Inc.}
  \city{San Francisco}
  \state{CA}
  \country{USA}}

\author{Raghav Saboo}
\authornotemark[1]
\email{raghav.saboo@doordash.com}
\affiliation{%
  \institution{DoorDash Inc.}
  \city{San Francisco}
  \state{CA}
  \country{USA}}

\author{Nimesh Sinha}
\email{nimesh.sinha@doordash.com}
\affiliation{%
  \institution{DoorDash Inc.}
  \city{San Francisco}
  \state{CA}
  \country{USA}}

\author{Leo Shao}
\email{leo.shao@doordash.com}
\affiliation{%
  \institution{DoorDash Inc.}
  \city{San Francisco}
  \state{CA}
  \country{USA}}

\author{Elyse Winer}
\email{elyse.winer@doordash.com}
\affiliation{%
  \institution{DoorDash Inc.}
  \city{San Francisco}
  \state{CA}
  \country{USA}}

\author{Sudeep Das}
\email{sudeep.das2@doordash.com}
\affiliation{%
  \institution{DoorDash Inc.}
  \city{San Francisco}
  \state{CA}
  \country{USA}}

\author{Martin Wang}
\email{martin.wang@doordash.com}
\affiliation{%
  \institution{DoorDash Inc.}
  \city{San Francisco}
  \state{CA}
  \country{USA}}

\author{Kyle MacDonald}
\email{kyle.macdonald@doordash.com}
\affiliation{%
  \institution{DoorDash Inc.}
  \city{San Francisco}
  \state{CA}
  \country{USA}}

\renewcommand{\shortauthors}{Xu et al.}

\begin{abstract}
Multi-merchant e-commerce catalogs contain equivalent and related products
under different merchant-scoped identifiers, fragmenting behavioral evidence
across merchants. Expert-defined taxonomies, meanwhile, are often too coarse
for fine-grained discovery. We investigate whether a single hierarchical
Semantic ID (\sid{}) representation can support personalized ranking and query
reformulation. Learned once from product-content embeddings, the hierarchy
defines product concepts at multiple granularities that each application
combines with its own behavioral and serving context. For ranking, we aggregate consumer affinity and product performance over
\sid{} prefixes and derive sequence features for candidate products and
consumer histories. Controlled ablations show improved offline relevance,
while online evaluation of the full ranking treatment shows stronger top-slot
add-to-cart engagement and broader exposure for less-popular products. For
query reformulation, we ground queries and session transitions in \sid{}
concepts, use the hierarchy for navigation and refinement, and filter
suggestions against the merchant's assortment. Offline evaluation shows finer
intent preservation than taxonomy and higher-quality suggestions than raw
query-string transitions; online evaluation shows reduced search effort and
earlier access to purchasable products. These results show that a shared
semantic product hierarchy can support both recommendation and search while
preserving the task-specific context required by each application.
\end{abstract}

\ccsdesc[500]{Information systems~Recommender systems}
\ccsdesc[500]{Information systems~Information retrieval}
\keywords{semantic IDs, hierarchical clustering, residual quantization, recommender systems, query reformulation}

\maketitle

\section{Introduction}
\label{sec:introduction}

E-commerce marketplaces bring together the catalogs of grocery, convenience,
and general retail merchants within a shared discovery experience. Equivalent
and closely related products often recur across merchants under different
listing IDs. A product representation must therefore capture relationships
across listings while preserving the distinctions that matter to consumers.
The representation used to organize products determines both the granularity
at which behavioral evidence can be aggregated and the products across which
that evidence can be shared.

This representation choice is particularly important for personalized ranking.
A merchant-scoped listing ID preserves exact identity, but an interaction with
a product at one merchant does not strengthen the preference signal for an
equivalent product elsewhere. A consumer may therefore purchase the same
product concept across multiple merchants without generating enough evidence
on any individual listing to learn a reliable preference. Human-defined
taxonomies share evidence across broader product groups, but their emphasis on
interpretability and catalog organization can make them too coarse for
fine-grained personalization.

A related representation problem arises in query reformulation, where candidate
reformulations are often derived from transitions between query strings
observed in search sessions. String-level transitions fragment evidence across
misspellings, abbreviations, and synonymous expressions. They can also be
dominated by broad, high-frequency queries and conflate distinct intents when
the same query has different meanings across business verticals. Products
associated with a query through downstream behavior provide a natural signal
for grounding its meaning in the catalog, allowing transitions to be modeled
between product concepts rather than raw strings. This requires a product
representation that is fine-grained enough to preserve useful distinctions and
hierarchical enough to support both lateral pivots, such as \emph{milk} to
\emph{cereal}, and narrower refinements, such as \emph{milk} to
\emph{whole milk}.

Semantic IDs (\sid{}s), introduced in TIGER~\cite{rajput2023tiger} for
generative retrieval, encode products as short sequences of discrete codes
obtained through residual quantization of product-content embeddings. Learned
\sid{}s often exhibit a hierarchical structure: products sharing a prefix tend
to be semantically related, with longer shared prefixes corresponding to
progressively finer-grained product groups. This structure provides multiple
levels at which related products can be grouped while retaining fine-grained
product distinctions.

\begin{figure*}[t]
  \centering
  \includegraphics[width=\textwidth]{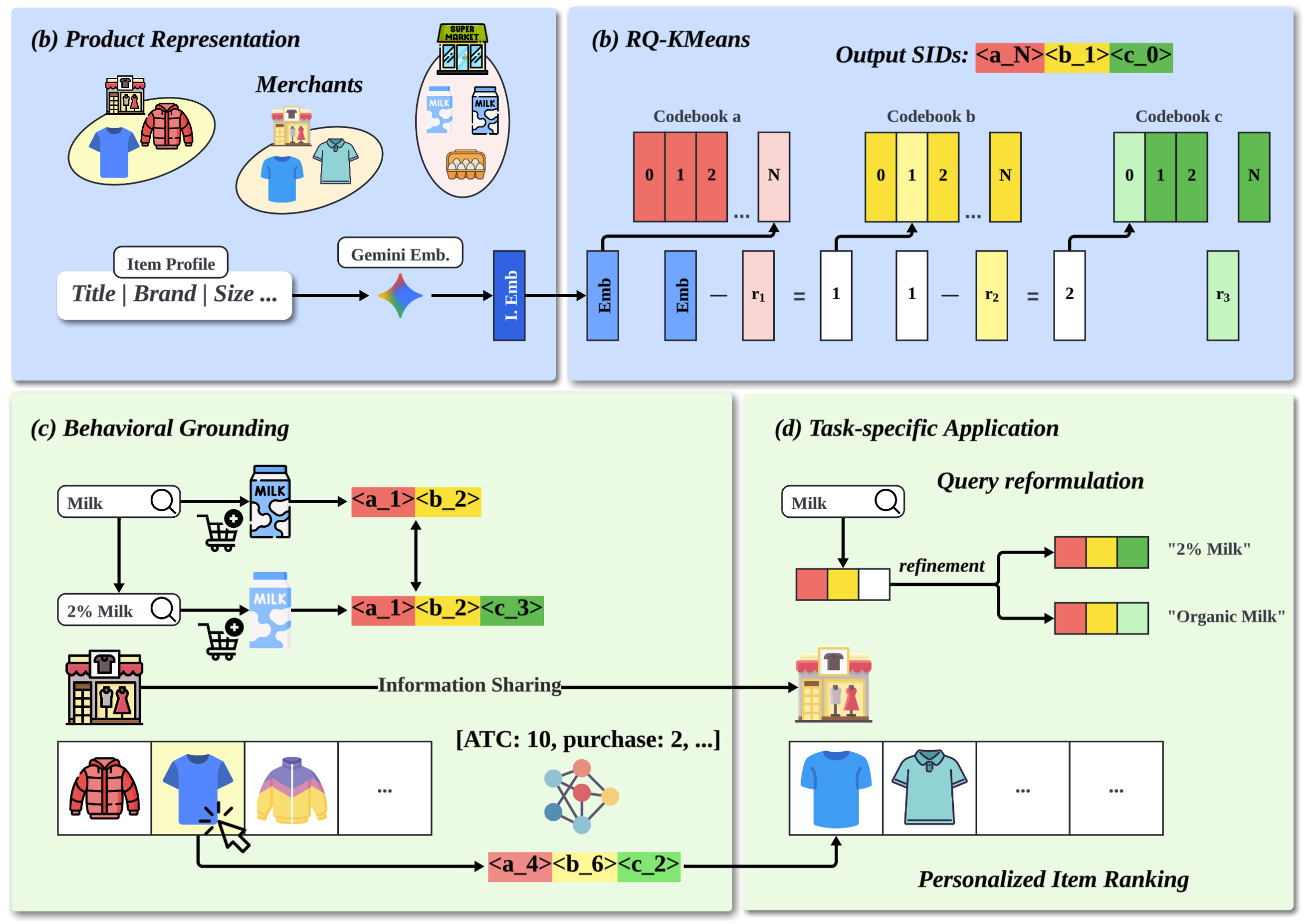}
  \caption{Overview of shared \sid{} construction and reuse across search and
    recommendation. Catalog product profiles are embedded and residual-quantized
    into a three-level \sid{} hierarchy. Search queries and consumer interactions
    are then grounded and aggregated through \sid{} prefixes, enabling
    cross-merchant preference transfer for personalized ranking and coarse-to-fine
    query reformulation through hierarchical descent.}
  \Description{A shared construction spine transforms product text into a
  three-level Semantic ID and two feature families. It branches into a
  personalized-ranking lane showing cross-merchant evidence transfer and a
  query-reformulation lane showing grounding, navigation, hierarchical descent,
  business-kind context, and merchant assortment.}
  \label{fig:overview}
\end{figure*}

In this paper, we propose using the learned hierarchy of \sid{}s as a shared
product representation across personalized ranking and query reformulation.
The hierarchy is constructed once from catalog content, while each application
independently organizes its interaction data around the product concepts it
defines. In personalized ranking, \sid{} prefixes serve as keys for behavioral
aggregates and as inputs to learned history representations, allowing
preferences to be modeled at multiple semantic resolutions. In query
reformulation, we map queries to \sid{} concepts using associated products,
model transitions between those concepts, and descend the hierarchy when a
more specific reformulation is appropriate. The two systems use different
levels of the hierarchy according to their respective objectives and do not share
model parameters, training objectives, or serving architecture.
Figure~\ref{fig:overview} summarizes the overall design.

Our contributions are threefold: (1) we present a production-scale study of
the learned hierarchy of \sid{}s as a reusable product representation across
recommendation and search; (2) we develop task-specific uses of this
representation in two independently developed applications: personalized item
ranking, where consumer interactions are aggregated over \sid{} prefixes at
multiple levels of granularity, and query reformulation, where the \sid{}
hierarchy grounds query transitions in product concepts and supports
hierarchical refinement; and (3) we provide empirical evidence that \sid{}s
offer a more effective representation than expert-defined taxonomy in both
applications, improving personalized ranking and enabling more specific,
assortment-aware query reformulations.

\section{Related Work}
\label{sec:related}

\paragraph{Semantic IDs.}
Semantic IDs (\sid{}s) map items to short sequences of discrete codes that
preserve structure from an underlying embedding space. Originally proposed in
TIGER~\cite{rajput2023tiger}, \sid{}s are constructed through residual
quantization of item embeddings and used as autoregressive targets for
next-item retrieval. Much of the subsequent work continues to study \sid{}s in
generative recommendation, search, and unified models of both
\cite{ju2025handbook,penha2025joint}. In industrial ranking, Singh et
al.~\cite{singh2023better} learn \sid{} subpieces using SentencePiece and use
their embeddings to represent items and user histories, finding them more
effective than manually defined (n)-grams. Zheng et
al.~\cite{zheng2025enhancing} develop prefix (n)-gram parameterizations and
deploy \sid{}-based sparse and sequential features for ads ranking. We
similarly use SentencePiece-tokenized \sid{}s to represent products and
consumer histories. Whereas prior ranking work primarily uses
\sid{}-derived tokens to parameterize learned representations, we additionally
use \sid{} prefixes as shared concept units across both applications:
personalized ranking aggregates consumer interactions over these concepts,
while query reformulation aggregates query-grounding and transition evidence
over the same hierarchy.

\paragraph{Query suggestion and reformulation.}
Behavioral query suggestion commonly represents queries as nodes connected by
transitions observed in search sessions. The Query-Flow
Graph~\cite{boldi2008queryflow} applies random walks over these transitions to
identify useful successor queries, while context-aware methods incorporate
click-through evidence to reduce sparsity and ambiguity
\cite{cao2008context}. Applied e-commerce systems also transfer behavioral
evidence from frequent queries to semantically related tail
queries~\cite{zhang2024reformulation}. Our approach retains the
interpretability and batch-serving advantages of a transition graph, but
grounds queries in product concepts and models transitions between concepts in
the \sid{} hierarchy. This allows evidence to be pooled across lexical
variants, supports both lateral reformulations and descent to more specific
concepts, and filters candidates against the merchant's active assortment.

\section{A Shared, Learned Product Hierarchy}
\label{sec:shared-sid}

\paragraph{Semantic ID construction.}
For product \(i\), we concatenate selected catalog fields, such as item name,
brand, and size, into a textual profile \(t_i\). We encode \(t_i\) using a
pretrained text encoder. In practice, we use
\texttt{gemini-embedding-001}~\cite{google2026geminiembedding}, which produces
a 3,072-dimensional embedding \(\mathbf{x}_i\).

We construct the \sid{} using residual-quantization \(K\)-means
(RQ-\(K\)-means) with \(L=3\) stages and \(K=512\) centroids per stage. Each
stage operates on an \(L_2\)-normalized input. Starting with \(\mathbf{r}_{i,0}=\mathbf{x}_i\), stage
\(\ell\in\{0,\ldots,L-1\}\) assigns the current residual to its nearest
centroid in codebook \(\mathcal{C}_{\ell}\) and subtracts the selected
centroid:
\[
\begin{aligned}
c_{i,\ell}
&=
\arg\min_{k\in\{0,\ldots,K-1\}}
\left\|
\mathbf{r}_{i,\ell}-\boldsymbol{\mu}_{\ell,k}
\right\|_2^2,\\
\widetilde{\mathbf{r}}_{i,\ell+1}
&=
\mathbf{r}_{i,\ell}
-
\boldsymbol{\mu}_{\ell,c_{i,\ell}}.
\end{aligned}
\]

The resulting SID is
\(\mathbf{s}_i=[c_{i,0},c_{i,1},c_{i,2}]\). Let
\(\mathbf{s}_i^{(\ell)}=[c_{i,0},\ldots,c_{i,\ell-1}]\) denote its prefix at
depth \(\ell\); we refer to depths \(1\), \(2\), and \(3\) as L1, L2, and L3,
respectively.

\paragraph{Learned hierarchy.}
The prefix \(n\)-grams form nested partitions of the product catalog. In the
learned code space, we observe that products sharing longer prefixes tend to
be more semantically similar, suggesting that the hierarchy captures product
concepts at increasing levels of specificity. These concepts provide an
intermediate granularity between exact listing IDs and expert-defined taxonomy
nodes. We use the same hierarchy in both applications: personalized ranking
aggregates consumer interactions over prefixes at multiple depths, while query
reformulation grounds queries and models transitions over the resulting
product concepts.

\paragraph{Characteristics of the learned hierarchy.}
We examine how the product groups induced by \sid{} prefixes change with
shared-prefix length. Table~\ref{tab:sid-diagnostics} reports code-usage
balance, cluster separation, and held-out semantic coherence for each prefix length.

\begin{table}[t]
  \centering
  \caption{Intrinsic characteristics of product groups induced by SID prefix
  levels. Lower Gini and DBI indicate more balanced code usage and
  better-separated groups, respectively; higher held-out cosine indicates
  stronger semantic coherence on unseen products.}
  \label{tab:sid-diagnostics}
  \small
  \begin{tabular}{@{}lccc@{}}
    \toprule
    Prefix level & Gini & DBI & Held-out cosine \\
    \midrule
    L1 (coarse)       & 0.472 & 3.934 & 0.957 \\
    L2 (intermediate) & 0.508 & 2.065 & 0.965 \\
    L3 (fine)         & 0.467 & 0.976 & 0.981 \\
    \bottomrule
  \end{tabular}
\end{table}

Gini measures imbalance in prefix usage, with lower values indicating a more
even distribution of products across groups. The Davies--Bouldin index (DBI)
compares within-group dispersion with between-group separation, with lower
values indicating more compact and better-separated groups. Held-out cosine
measures the average cosine similarity between an unseen product and training
products assigned to the same prefix group.

As shared-prefix length increases, DBI decreases and held-out cosine
similarity increases, indicating that longer prefixes identify more compact
and semantically coherent product groups. Code-usage imbalance remains
comparable across the three prefix lengths. These observations suggest that
the learned \sid{} hierarchy captures product concepts at progressively finer
levels of granularity.

\section{Personalized Item Ranking with Semantic IDs}
\label{sec:ranking}

We first examine how the learned \sid{} hierarchy supports personalized item
ranking on discovery surfaces. On these surfaces, consumers browse a
merchant's assortment through category- and theme-based carousels. A carousel
may represent a grocery mission such as \emph{Produce} or
\emph{Summer Grilling}, or a retail concept such as \emph{Beauty},
\emph{Apparel}, or \emph{Household Supplies}. A retrieval stage selects
candidate items for each carousel, and a personalized ranker determines their
order for each consumer.

\subsection{Ranking Model}

The ranker uses a multi-task, multi-label neural network with CTR, ATCR, and
CVR heads, corresponding to click, add-to-cart, and purchase outcomes,
respectively. Its existing
product-identity features include merchant-scoped listing IDs and
expert-defined taxonomy prefixes. Listing IDs preserve exact identity within a
merchant but cannot transfer a consumer's history to equivalent or related
products at another merchant. Taxonomy prefixes support broader sharing but
may group products that are too heterogeneous to represent a specific grocery
or retail preference. Consequently, established fine-grained preferences may
not be reflected near the top of a carousel, particularly at merchants from
which the consumer has not previously ordered.

\subsection{Feature Design}

We construct two complementary families of \sid{}-derived features. Dense
features explicitly aggregate behavioral statistics over the product concepts
defined by \sid{} prefixes. This provides a strong inductive bias without
requiring the ranker to learn each high-cardinality concept solely from a
trainable embedding. Sequence features preserve information from the \sid{}
code sequence and support learned interactions between candidate products and
consumer histories. Table~\ref{tab:rankfeat} summarizes representative features from both
families.

Following prior work on \sid{} parameterization
\cite{singh2023better,zheng2025enhancing}, we use the prefix \(n\)-grams
defined in Section~\ref{sec:shared-sid} as aggregation keys for dense features
and SentencePiece subwords as the units of item- and consumer-side sequence
features.

\paragraph{Feature identifiers.}
Each prefix \(n\)-gram is mapped to a deterministic integer identifier:
\[
  p_i^{(n)}
  =
  \sum_{j=0}^{n-1}
  \left((c_{i,j}+1)K^j-1\right),
  \qquad n\in\{1,2,3\},\quad K=512.
\]
The identifiers for L1, L2, and L3 are stored in separate fields, making the
mapping collision-free within each prefix level.

To construct SentencePiece inputs, we map each
\((\text{position},\text{code})\) pair to a distinct symbol from an alphabet
of \(3K=1{,}536\) symbols, converting each three-code \sid{} into a
three-symbol string. We train SentencePiece models
\cite{kudo2018sentencepiece} on an impression-weighted corpus of these
strings. The resulting pieces may span one, two, or three adjacent \sid{}
symbols and therefore need not coincide with hierarchy prefixes.

\begin{table*}[t]
  \centering
  \caption{Examples of \sid{}-derived ranking features.}
  \label{tab:rankfeat}
  \small
  \renewcommand{\arraystretch}{1.15}
  \begin{tabular}{
    @{}
    >{\raggedright\arraybackslash}p{0.16\textwidth}
    >{\raggedright\arraybackslash}p{0.18\textwidth}
    >{\raggedright\arraybackslash}p{0.58\textwidth}
    @{}
  }
    \toprule
    \textbf{Feature family} & \textbf{Feature Scope} & \textbf{Description} \\
    \midrule

    \multirow{2}{*}{Dense aggregates}
    & Consumer
    & Order frequency, purchase recency, and subtotal statistics for each
      \sid{} prefix in the consumer's history. \\

    & Submarket
    & Impressions, clicks, \atc{} actions, purchases, and associated rates for
      each \sid{} prefix within a submarket. \\

    \midrule

    \multirow{2}{*}{Sequence}
    & Item
    & SPM token sequence obtained from the candidate item's \sid{}. \\

    & Consumer
    & SPM tokens aggregated from previously ordered products and ranked by
      their associated order counts. \\

    \bottomrule
  \end{tabular}
\end{table*}

\paragraph{Dense aggregate features.}
\sid{} prefixes allow behavioral evidence to be shared at several semantic
resolutions. Consumer-level aggregates capture individual affinity for the
product concepts represented by each prefix, while global and regional
aggregates provide overall and locally conditioned performance priors. We
compute these signals at all three prefix levels and, where applicable, over
multiple lookback windows. The ranker can therefore combine the broader
coverage of shallow prefixes with the greater specificity of deeper prefixes.

\paragraph{Sequence features.}
We evaluated both the Unigram LM and BPE variants of SentencePiece. The
Unigram LM vocabulary consisted largely of \sid{} prefix \(n\)-grams and
therefore overlapped substantially with the information already captured by
the prefix-keyed dense features. We use BPE to obtain a more complementary set
of subsequences and refer to the resulting units as SPM tokens.

On the item side, the candidate product is represented by the SPM tokens
obtained from its \sid{}. On the consumer side, we collect products ordered
during the preceding 180 days, expand their SPM token lists, and aggregate
order frequency by token. Distinct tokens are ranked by order count, with
recency used as a tie-breaker, and the highest-ranked tokens are retained as a
compact representation of recurring semantic preferences.

The item- and consumer-side tokens share an embedding table, allowing the
ranker to relate candidate-product subwords directly to the consumer's
historical preferences. Let \(\mathcal{T}\) denote the SPM token list for
either input. We use a vocabulary of \(N=2\times10^5\) learned tokens and
reserve one additional entry for the null token, giving a shared trainable
embedding table
\(\mathbf{E}\in\mathbb{R}^{(N+1)\times64}\). We obtain the representation of
each input by mean pooling:
\[
\mathbf{h}(\mathcal{T})
=
\frac{1}{|\mathcal{T}|}
\sum_{t\in\mathcal{T}}\mathbf{E}_{t}.
\]
A separate mask indicates whether the input is empty. The pooled
representations and embedding parameters are learned jointly with the ranking
objective.

\subsection{Offline Evaluation}
\label{subsec:ranking-offline}

The ranking model is trained on \(M\) days of logged interactions and evaluated
on data from day \(M+1\). The full candidate (FC) includes \sid{}-derived
features alongside concurrent non-\sid{} feature updates, so comparison with
the production baseline alone would not isolate the contribution of \sid{}. We
therefore construct an ablated candidate (FC-A) that removes all
\sid{}-derived features while holding the complete non-\sid{} feature
configuration fixed.

We evaluate FC and FC-A on the CVR head using MRR@\(K\) and NDCG@\(K\) for
\(K\in\{3,5,10\}\). We report \(K=5\) in Table~\ref{tab:ranking-offline}
because five items are visible on the evaluated discovery surface without
scrolling. Both metrics are computed over sessions containing at least one
conversion.

\begin{table}[t]
  \centering
  \caption{Offline ablation of \sid{}-derived ranking features. All values are
  relative gains over the production baseline.}
  \label{tab:ranking-offline}
  \small
  \begin{tabular}{@{}lrr@{}}
    \toprule
    Model & MRR@5 & NDCG@5 \\
    \midrule
    FC-A & \(+2.10\%\) & \(+2.92\%\) \\
    FC   & \(+6.98\%\) & \(+6.76\%\) \\
    \bottomrule
  \end{tabular}
\end{table}

As shown in Table~\ref{tab:ranking-offline}, FC substantially outperforms FC-A
on both reported metrics. The same pattern holds at \(K=3\) and \(K=10\).
Since the two candidates differ only in their \sid{}-derived features, the
ablation indicates that these features contribute the majority of the full
candidate's offline gain.

\subsection{Online Experiment}
\label{subsec:ranking-online}

The online experiment evaluates a production feature bundle that combines the
\sid{}-derived features with concurrent non-\sid{} updates. Consumers were
randomized approximately evenly among three arms: the existing ranker, the
feature-bundle treatment, and the same treatment with an additional serving
optimization. We report the 21-day comparison between the existing ranker and
the feature-bundle treatment.

\begin{table}[t]
  \centering
  \caption{Online results for the full ranking treatment. The treatment
  includes \sid{}-derived features alongside concurrent non-\sid{} updates.}
  \label{tab:ranking-online}
  \small
  \begin{tabular}{@{}lr@{}}
    \toprule
    Outcome & Relative gain \\
    \midrule
    Subtotal
      & \(+0.31\%\) \\
    Average carousel \atc{} rate
      & \(+5.5\%\) \\
    Item \atc{} rate, position 1
      & \(+8\%\) \\
    Item \atc{} rate, position 2
      & \(+16\%\) \\
    Item \atc{} rate, position 3
      & \(+6\%\) \\
    \bottomrule
  \end{tabular}
\end{table}

As shown in Table~\ref{tab:ranking-online}, the treatment improves
add-to-cart engagement across the carousel and at each of the first three
positions. These engagement gains translate into a \(0.31\%\) relative
increase in subtotal.

The treatment also reduces popularity concentration. The average historical
popularity of the item displayed in the first carousel position decreases by
\(18.1\%\), while the share of first-position impressions assigned to
blockbuster items decreases by \(2.1\) percentage points. Thus, the treatment
improves engagement while allocating less top-position exposure to
historically dominant products.

Because the production bundle includes both \sid{} and non-\sid{} updates, the
online experiment alone does not isolate the contribution of \sid{}.
However, both the offline ablation study and post-experiment analysis indicate
that \sid{}-derived features account for most of the observed improvement.

\begin{table}[t]
  \centering
  \caption{Online impact of \sid{}-based query reformulation on search efficiency. Changes are relative to control; lower
\atc{} position and scroll depth are better.}
  \label{tab:query-online}
  \small
  \begin{tabular}{@{}lrr@{}}
    \toprule
    Metric & Relative change & 95\% CI \\
    \midrule
    Purchase MRR
      & \(+0.558\%\) & \([+0.294,+0.823]\%\) \\
    \atc{} position
      & \(-1.571\%\) & \([-2.159,-0.982]\%\) \\
    Search scroll depth
      & \(-1.866\%\) & \([-2.262,-1.470]\%\) \\
    \bottomrule
  \end{tabular}
\end{table}

\section{Query Reformulation with Semantic IDs}
\label{sec:query}

Beyond personalized item ranking, the same \sid{} representation can support
search-time intent discovery. We study query reformulation for suggested-query pills displayed alongside
search results on a merchant's Store Page. A useful suggestion must advance the
consumer's shopping mission and correspond to products in the merchant's
active assortment. Shopping sessions may involve lateral basket-building moves,
such as \emph{milk} to \emph{cereal}, or refinements to a product type, brand,
or variant.

Prior work on query reformulation often mines query-to-query transitions directly from search
sessions. Representing queries as raw strings fragments behavioral evidence
across misspellings, abbreviations, and synonymous expressions, while pooling
the same string across business kinds can conflate different meanings. We
instead use the \sid{} hierarchy as a catalog-grounded concept space that
supports both lateral navigation and progressive refinement.

\subsection{Methodology}
\label{subsec:query-method}

\paragraph{Query-to-concept grounding.}
For each query and business vertical (BV), we aggregate associated \atc{}
events by \sid{} prefix. For a query--BV pair with at least five events, we
assign the dominant L2 prefix if it accounts for at least \(30\%\) of the
evidence. Otherwise, we apply the same criterion at L1. If neither level
qualifies, a guarded fragmented-query path retains the raw query node rather
than assigning an unsupported concept. Conditioning on BV allows the same
query string to resolve to different product concepts across retail contexts.

\paragraph{Lateral navigation.}
Given consecutive queries \((q_t,q_{t+1})\) in a session, we replace the
string pair with a transition between their grounded \sid{} prefixes. We
estimate transition counts and marginals separately within each BV and rank
candidate edges by normalized pointwise mutual information
(NPMI)~\cite{bouma2009npmi}. We retain an edge when its observed count is at
least \(5\), its expected count under independence is at least \(1\), and its
NPMI is at least \(0.1\). This construction pools lexical variants that resolve
to the same concept while preserving BV-specific interpretations of ambiguous
queries.

\paragraph{Hierarchical refinement.}
Queries grounded to the same \sid{} prefix share a graph node, causing
transitions between finer-grained intents to collapse into self-loops and be
discarded. This is particularly limiting at L2, where the graph captures
lateral pivots across concepts but misses refinements within a shared parent.
We therefore add a parallel path that descends from an L2 prefix to its L3
children, ranking them with query-specific \atc{} evidence when available and
parent-level popularity otherwise.

\paragraph{Query rendering.}
Because \sid{}s are internal identifiers, a language-model prompt renders
target concepts from both candidate-generation paths as short consumer-facing
queries using representative products from each concept. A second pass removes
unusable source--target pairs, and the remaining candidates are ranked by
embedding similarity between the source query and the rendered query.
Language generation is confined to this rendering step; observed behavior and
catalog concepts determine the candidate structure.

\paragraph{Assortment-aware filtering.}
At serving time, candidates are filtered against the merchant's active
assortment. A target concept is eligible when the assortment contains at least
one active item assigned to that concept, preventing suggestions for intents
the merchant cannot fulfill.

\subsection{Offline Evaluation}
\label{subsec:qr-eval}

Our offline evaluation focuses on two complementary questions: whether
\sid{}s preserve fine-grained intent distinctions better than the product
taxonomy, and whether transitions over \sid{} concepts produce better
reformulations than transitions over raw query strings.

For the taxonomy comparison, we measure how often queries expressing different
intents map to the same concept. Such transitions collapse into self-loops and
cannot generate reformulation candidates. Taxonomy collapses \(18.8\%\) of
intent-changing transitions, compared with \(10.9\%\) for \sid{}s. The finer
\sid{} representation therefore preserves more observed behavioral signal for
candidate generation.

We next compare end-to-end suggestion quality against a query-string
transition graph. An LLM judge scores usefulness, target-text quality, and
distinctiveness using the labels \emph{bad}, \emph{acceptable}, and
\emph{good}, which are mapped to \(0\), \(0.5\), and \(1\), respectively.
On a human-labeled set of 200 query pairs, the judge achieves \(78\%\) exact
agreement and \(80\%\) agreement on usefulness. Among queries served by both
systems, rank-one judged quality increases from \(0.522\) for the query-string
graph to \(0.734\) for the catalog-grounded \sid{} system.

Together, these results indicate that \sid{}s provide a more discriminative
representation of product intent than taxonomy and a stronger foundation for
mining behavioral query transitions than raw query strings.

\subsection{Online Experiment}
\label{subsec:qr-online}

We evaluate the complete \sid{}-based reformulation module in a
consumer-randomized experiment against a control without suggested-query
reformulations. Table~\ref{tab:query-online} reports selected outcomes as
relative changes from control.

The reformulation module improves purchase MRR while reducing both the
position of ATC and search scroll depth. Together, these
changes indicate that consumers reach relevant, purchasable items earlier and
with less search effort.

\section{Qualitative Analysis}
\label{sec:qualitative}

Appendix Table~\ref{tab:examples} illustrates two properties of the \sid{}
representation that benefit both applications. First, products with similar
semantics share codes even when they belong to different merchant-scoped
listings, allowing behavioral evidence to transfer across merchants. Second,
the prefix hierarchy exposes multiple levels of granularity: coarser prefixes
pool evidence, while deeper prefixes distinguish more specific product intents.

These examples also expose the trade-off introduced by semantic compression.
Mapping many products or queries to one prefix increases statistical support,
but an overly broad or semantically mixed prefix can connect unrelated
behaviors. In reformulation, such prefixes may become high-degree graph hubs
that propagate irrelevant suggestions. Product-based query grounding can also
capture the concept ultimately added to cart rather than the intent expressed
by the original query. In ranking, products near a quantization boundary may
receive different codes despite being useful substitutes, while products
sharing a prefix may still differ on attributes that matter to a particular
consumer.

The applications address this trade-off by restoring information outside the
\sid{} itself. Ranking combines prefixes at multiple depths with
consumer-specific and product-specific signals. Query reformulation conditions
grounding and transitions on BV, uses query-specific evidence when descending
to L3, and filters candidates against the merchant's assortment. The \sid{}
hierarchy therefore supplies a transferable semantic prior rather than a
complete representation of task intent.

\section{Cross-System Findings}
\label{sec:findings}

Appendix Table~\ref{tab:cross-system} summarizes how the two applications use the same
\sid{} hierarchy. In both cases, \sid{}s replace a fragmented behavioral unit
with a semantic unit over which evidence can be pooled. The applications differ
in which hierarchy depths they use, which behavioral signals they attach, and
which information they restore before producing an output.

The shared pattern is to use \sid{}s for semantic evidence pooling and recover
task-specific context before acting. Coarser prefixes provide support and
transferability, while deeper prefixes recover specificity. The appropriate
operating depth is application-dependent: ranking can consume several depths
jointly, whereas reformulation assigns distinct roles to L2 navigation and L3
refinement.

This separation also clarifies the role of \sid{}s relative to existing
identifiers. Exact product IDs remain necessary for identity and serving, and
taxonomy remains useful for business organization. \sid{}s complement these
representations by supplying a fine-grained, transferable hierarchy that can be
reused across discovery tasks without requiring the applications to share the
same model or decision logic.

\section{Conclusion}
\label{sec:conclusion}

We show that a single hierarchical Semantic ID vocabulary can support
discovery systems spanning recommendation and search. Personalized item
ranking uses \sid{} prefixes as transferable feature keys over
consumer--product interactions, while query reformulation uses the same
hierarchy to ground queries, mine concept transitions, and generate
coarse-to-fine suggestions within a merchant's assortment.

Across both applications, \sid{}s improve the task-specific outcomes enabled
by the shared hierarchy. In ranking, \sid{}-derived features add incremental
predictive signal and contribute materially to relevance gains. In query
reformulation, \sid{} concepts preserve finer intent distinctions than
taxonomy, improve suggestion quality over query-string transitions, and reduce
search effort online. These findings show that search and recommendation can
share a semantic product representation without requiring a jointly trained
model. \sid{}s complement exact identifiers and product taxonomies with a
fine-grained, transferable hierarchy that each application can combine with
its own behavioral signals and serving context.

\balance
\begin{acks}
We thank Kevin Zhai, Abhishek Tambat, Camrick Solorio, Akshad Viswanathan,
Danielle Rommerdahl, Natalia Dougan, Vivek Paharia, Kyle Hsiao, and
Tito Anammah for their support in this work.
\end{acks}

\bibliographystyle{ACM-Reference-Format}
\bibliography{refs}

\clearpage
\appendix
\onecolumn

\section{Qualitative Examples and Cross-System Comparison}
\label{app:cross-system}
\begin{table*}[h]
  \centering
  \caption{Illustrative effects of \sid{} transferability and hierarchical
  granularity.}
  \label{tab:examples}
  \small
  \setlength{\tabcolsep}{6pt}
  \renewcommand{\arraystretch}{1.12}

  \begin{tabularx}{\textwidth}{@{}L{0.21\textwidth}YY@{}}
    \toprule
    \sid{} property & Example & Observed effect \\
    \midrule

    \multicolumn{3}{@{}l}{\textbf{Personalized item ranking}} \\
    \addlinespace[2pt]

    Cross-merchant transfer
      & Prior purchases of mushrooms and chicken at another merchant
      & Matching products rise toward the top despite having different
        merchant-scoped identifiers. \\
    \addlinespace[3pt]

    Fine-grained semantic affinity
      & Prior purchase of a makeup brush
      & A related brush set rises above a makeup sponge, and more brush
        products appear among the top results. \\

    \midrule
    \multicolumn{3}{@{}l}{\textbf{Query reformulation}} \\
    \addlinespace[2pt]

    Hierarchical refinement
      & Source query: \emph{clay mask}
      & L3 descent introduces more specific intents, including
        \emph{sheet mask} and \emph{hydrating mask}. \\
    \addlinespace[3pt]

    Coarse-to-fine navigation
      & Source query: \emph{car clean}
      & L3 adds \emph{interior wipes}, \emph{interior cleaner}, and
        \emph{leather wipes}, while L2 retains a useful lateral pivot. \\

    \bottomrule
  \end{tabularx}
\end{table*}

\begin{table*}[h]
  \centering
  \caption{How personalized ranking and query reformulation reuse the shared
  \sid{} hierarchy.}
  \label{tab:cross-system}
  \small
  \setlength{\tabcolsep}{6pt}
  \renewcommand{\arraystretch}{1.12}

  \begin{tabularx}{\textwidth}{@{}L{0.18\textwidth}YY@{}}
    \toprule
    \sid{} role
      & Personalized item ranking
      & Query reformulation \\
    \midrule

    Input unit
      & Merchant-scoped product listings
      & Raw query strings \\
    \addlinespace[2pt]

    Mapping into \sid{} space
      & Products in consumer interaction histories
      & Queries grounded through associated products \\
    \addlinespace[2pt]

    Evidence pooled
      & Cross-merchant interactions among semantically related products
      & Transitions across lexical variants of the same concept \\
    \addlinespace[2pt]

    Hierarchy use
      & L1--L3 prefixes used jointly as ranking features
      & L2 transitions for navigation; L3 descent for refinement \\
    \addlinespace[2pt]

    Task-specific signals
      & Affinity, recency, product performance, and interaction sequences
      & Transition strength, query evidence, and concept popularity \\
    \addlinespace[2pt]

    Context restored
      & Consumer, candidate-product, and regional context
      & BV, source query, and merchant assortment \\
    \addlinespace[2pt]

    Main representation risk
      & Shared prefixes may hide preference-relevant attributes
      & Broad prefixes may collapse intents or become graph hubs \\

    \bottomrule
  \end{tabularx}
\end{table*}

\end{document}